\documentclass[proof]{WileyASNA-v1}
\usepackage{orcidlink} % ORCID 
\usepackage{hyperref} % Hyperlink 
\articletype{Article Type}%

\received{XX XX XXXX}
\revised{XX XX XXXX}
\accepted{XX XX XXXX}

\begin{document}

\title{Enhanced kinematics and distribution characteristics of Upper Scorpius and Ophiuchus associations based on Gaia DR3}

\author[1,2]{W. H. Elsanhoury*}
%\orcidlink{0000-0002-2298-4026}

%\author[1,2]{W. H. Elsanhoury* \hyperlink{https://orcid.org/0000-0002-2298-4026}{https://orcid.org/0000-0002-2298-4026}}

\address[1]{\orgdiv{Physics Department, College of Science}, \orgname{Northern Border University}, \orgaddress{\state{Arar}, \country{Saudi Arabia}}}

\address[2]{\orgdiv{Astronomy Department}, \orgname{National Research Institute of Astronomy and Geophysics (NRIAG)}, \orgaddress{\state{11421, Helwan, Cairo}, \country{Egypt}}}

\corres{*W. H. Elsanhoury, Physics Department, College of Science, Northern Border University, Arar, Saudi Arabia.
\\
\email{elsanhoury@nriag.sci.eg} $\&$ {elsanhoury@nbu.edu.sa}}

\presentaddress{Physics Department, College of Science, Northern Border University, Arar, Saudi Arabia}

\authormark{ELSANHOURY}

\abstract[Abstract]{The kinematics within the Solar vicinity have revealed interesting features relevant to both stellar and Galactic structures. This study examines three stellar associations in the Upper Scorpius and Ophiuchus regions, along with their sub-samples among Gaia DR3. The calculated kinematics and velocity ellipsoid characteristics, including the mean spatial velocity components
($U, ~V,$ and $W$; km s$^{-1}$), yielding values of approximately (-5.84 $\pm$ 2.42, -16.14 $\pm$ 4.02, -7.31 $\pm$ 2.70; km s$^{-1}$). USC and Oph associations velocity dispersion within the ellipsoid ($\sigma_i,~\forall{i=1,~2,~3}$) was found to be (1.36 $\pm$ 0.02, 0.80 $\pm$ 0.01, 0.96 $\pm$ 0.01; km s$^{-1}$), their mean Solar motion ($S_\odot$) was determined to be approximately 18.62 $\pm$ 4.32 km s$^{-1}$, convergent point coordinates ($A_o,~D_o$) were (95$^{o}$.91 $\pm$ 0$^{o}$.09, -44$^{o}$.42 $\pm$ 0$^{o}$.02), and Oort’s constants, yielding $A$ = 17.80 $\pm$ 0.24 km s$^{-1}$ kpc$^{-1}$ and $B$ = -9.61 $\pm$ 0.32 km s$^{-1}$ kpc$^{-1}$. Finally, the density distribution function per absolute magnitudes of USC and Oph associations is examined to obtain both luminosity and mass functions; their analysis revealed the absence of any peaks or dips as consistent with other recent studies.}


\maketitle

\section{Introduction}
\label{sec1}
The Scorpius-Centaurus OB association (Sco-Cen) is the nearest massive star formation vicinity to the Sun, spanning a large area in the sky due to its proximity \citep{2008hsf2.book..235P,2023MNRAS.519.3992Z}. \cite{1947esa..book.....A} coined the term “association” to describe these clusters of OB stars, highlighting that their stellar mass density often falls below 0.1 $M_{\odot}$ pc\textsuperscript{-3}. OB associations are relatively young as \cite{1934HarCi.384....1B} demonstrated that such low-density star clusters are unstable against Galactic tidal forces \citep{1949AZh....26....3A}. The Gaia mission's precision astrometry \citep{2016A&A...595A...1G,2018A&A...616A...1G,2021A&A...649A...1G} has allowed for a more detailed view of the association, resulting in a rise in the number of confirmed Sco-Cen members from just 512 stars \citep{1999AJ....117..354D} to nearly 15,000 candidate objects \citep{2019A&A...623A.112D}.

Three groupings have historically been identified within Sco-Cen: Upper Scorpius (USC), Upper Centaurus-Lupus (UCL), and Lower Centaurus-Crux (LCC) \citep{1964ARA&A...2..213B}. Numerous studies employing evolutionary models that incorporate various physical parameters indicate that the USC associations are indeed young $\sim$5-10 Myr \citep{2002AJ....124..404P,2016MNRAS.461..794P} and located at distance $d\sim$145 pc \citep{2008hsf2.book..235P}. The majority star formation studies, including those examining the mass function \citep{2022NatAs...6...89M} and disc and planet formation \citep{2016ApJ...827..142B,2018AJ....156...75E,2018MNRAS.477.5191R}, have been significantly influenced by this age uncertainty.

With a high concentration of protostars and cold gas, the Ophiuchus (Oph) molecular cloud (MC) is one of the densest areas of the Sco-Cen as part of the association. The region occupied by USC and Oph is one of the nearest star forming complexes, providing an ideal laboratory for studying the initiation and propagation of star formation through MCs \citep{2022NatAs...6...89M}. Although USC is located near the Oph star-forming clouds ($345^{o}\le l\le10^{o}$, $0^{o}\le b \le 25^{o}$), no signs of active star formation are currently observed. 

A comprehensive census of the USC association members requires wide-field imaging data due to its large spatial extent, covering approximately ($\sim$100 deg$^2$) of the sky. Given the low extinction in USC ($A_V<3$ mag), extensive field surveys have been conducted across a wide range of wavelengths to map the region (e.g., \cite{2006MNRAS.373...95L}; \cite{2006AJ....131.3016S}; \cite{2015MNRAS.448.2737R}; \cite{2016MNRAS.461..794P}). The USC association includes many of the bright stars in the constellations Scorpius (Sco), Lupus, Centaurus, and Crux, including Antares, the most massive star in USC, and the majority of the stars in the Southern Cross \citep{2008hsf2.book..235P}. The MCs in Oph region have already dissipated, allowing for the observation of cluster members spanning a wide range of masses. An examination of the association’s high-mass star population ($M\ge2M_\odot$) \citep{1999AJ....117..354D} determined that the majority of the group’s members exhibit spectral types $B$, $A$, and $F$.

The study aims to re-examine and report the spatial structures, kinematics of the USC and Oph complex, including their velocity ellipsoid motion characteristics, parameters characterizing the local rotational properties of our Galaxy, such as Oort’s constants $A$ and $B$, and their distribution among luminosity and mass functions. 

In the context of our ongoing investigations into stellar associations, we present velocity ellipsoid parameters (VEPs) for the USC and Oph associations, derived from Gaia data. Our previous studies have explored VEPs for various associations, including those of cool and ultra-cool stars \citep{2016Ap.....59..246E}, the kinematics and ellipsoidal motion of mid-to-late M-type stars \citep{2021AN....342..989E}, K dwarfs \citep{elsanhouryAN}, inner-halo hot sub-dwarfs \citep{elsanhoury24}, and for our first series, \citep{2015RMxAA..51..199E} its considered to compute the velocity ellipsoid of Solar neighborhood white dwarfs located $\sim$ 25 pc.

The remainder of this paper is structured as follows: Section \ref{sec2} provides an overview of the selected data. Section \ref{sec3} details the computational methods, including inner kinematics, convergent point analysis, Solar element determination, and the Oort constants. Section \ref{sec5} refers to luminosity and mass functions within USC and Oph associations. Finally, Section \ref{sec6} presents the discussion and conclusions.

\maketitle

\section{Data and sample selection}
\label{sec2}
The data used in this analysis is aimed to build a census of known members within both the USC and Oph associations. The Gaia mission, originally known as the Global Astrometric Interferometer for Astrophysics, is a project by the European Space Agency designed to create the most comprehensive and accurate three-dimensional map of our Galaxy. Gaia is conducting an unprecedented survey of one percent of the Galaxy’s 100 billion stars with microarcsecond ($\mu$as) precision.

The third data release (DR3) from Gaia mission \citep{2023A&A...674A...1G} (hereafter DR3) represents a major milestone in astronomy. DR3 includes five-parameter astrometry for approximately 1.8 billion sources, providing data on sky positions ($\alpha,~\delta$), parallaxes ($\pi$; mas), and proper motion components in right ascension and declination ($\mu_{\alpha}\cos{\delta},~\mu_{\delta}$; mas yr\textsuperscript{-1}), with a limiting magnitude of $G=21$ mag. The uncertainties in proper motion measurements vary: 0.02–0.03 mas yr\textsuperscript{-1} for sources with $G<15$ mag, 0.07 mas yr\textsuperscript{-1} for $G$ $\sim$ 17 mag, 0.50 mas yr\textsuperscript{-1} for $G$ $\sim$ 20 mag, and 1.40 mas yr\textsuperscript{-1} for $G=21$. Gaia parallax errors are $\sim$ 0.02–0.03 mas for $G<15$ mag, $\sim$ 0.07 mas for $G$ = 17 mag, $\sim$ 0.50 mas for $G$ = 20 mag, and $\sim$ 1.30 mas for $G$ = 21.

Radial velocities ($V_r$; km s\textsuperscript{-1}) for approximately 7 million stars from DR2 were also incorporated into the DR3 dataset \citep{2021A&A...649A...1G}. Notably, DR3 features improved astrometric accuracy, with a twofold improvement in proper motion accuracy and approximately a 1.5-fold improvement in parallax accuracy compared to DR2. Additionally, proper motion measurements experienced a 2.5-fold improvement, while parallax measurements experienced a 30–40$\%$ reduction in astrometric inaccuracies.

For this study, we focused on three specific Programs:

1) Program I: established by \citet{2018MNRAS.477L..50G}\footnote{\url{https://vizier.cds.unistra.fr/viz-bin/VizieR?-source=J/MNRAS/477/L50}}, compiled a list of 1322 likely member stars of the USC association, originally derived from the Two Micron All-Sky Survey (2MASS) \citep{2006AJ....131.1163S}. This selection was based on youth diagnostics and proper motions \citep{1999AJ....117..354D,1999AJ....117.2381P,2002AJ....124..404P,2011MNRAS.416.3108R,2012ApJ...758...31L,2015MNRAS.448.2737R,2016ApJ...827..142B}. Trigonometric parallaxes and proper motions were sourced from the Tycho–Gaia Astrometric Solution (TGAS) \citep{2016A&A...595A...4L}. After comparing the proper motions of stars in their sample, as provided by several catalogs, \cite{2018MNRAS.477L..50G} identified a few outliers in the fifth United States Naval Observatory CCD Astrograph Catalog (UCAC5) \citep{2013AJ....145...44Z}. Consequently, they supplemented the UCAC5 proper motions with data from the fourth Southern Proper Motion Catalog (SPM4) \citep{2011AJ....142...15G}, TGAS, and the Hot Stuff for One Year catalog \citep{2017A&A...600L...4A}, which served as their primary source of proper motions. Additionally, they used data mining tools from the Strasbourg Astronomical Data Center (CDS/SIMBAD) \citep{2000A&AS..143....9W} and other sources (e.g., \cite{2012AJ....144....8S,2013AJ....146..134K,2014ApJ...788...81M}; \cite{2016AJ....152...61M}) to search the available literature for information on radial velocity. In total, they discovered the radial velocities of 146 stars.

For the present study, the downloaded kinematic data of ($U,~V$, and $W$; km s\textsuperscript{-1}) was done from \cite{2018MNRAS.477L..50G} to create the worksheet data required for constructing our first Program of USC, followed by updated using the most recent data from \cite{2022yCat.1355....0G}. To ensure data consistency cross-match were carried out using software that was based on the Tool for OPerations on Catalogues And Tables (TOPCAT) and Starlink Tables Infrastructure Library (STIL) \citep{2005ASPC..347...29T}. This tool is particularly robust in analyzing tabular data within a specified range ($0\le x<1$) and offers numerous features for manipulating astronomical catalogs. Through this process, we identified 51 true members, as listed in Table \ref{tab1}. The table keys are organized as follows: the first column represents the equatorial coordinates ($\alpha,~\delta$)\textsubscript{2016}, the second and third columns contain 2MASS and Gaia DR3 identifier numbers, respectively, and the last three columns list the spatial velocities ($U,~V$, and $W$; km s$^{-1}$) with their uncertainties ($\sigma_U,~\sigma_V,$ and $\sigma_W$; km s$^{-1}$) relative to the Galactic coordinates system.

\begin{center}
\begin{table*}[t]%
\caption{Program I: Spatial velocities for 51 USC association \citep{2018MNRAS.477L..50G}}
\label{tab1}
\centering
\begin{tabular*}{500pt}{@{\extracolsep\fill}lcccccD{.}{.}{3}c@{\extracolsep\fill}}
\toprule

($\alpha,~\delta$)$^{o}_{2016}$ & 2MASS  & Gaia DR3  & $U\pm\sigma_{U}$  & $V\pm\sigma_{V}$ & $W\pm\sigma_{W}$ \\
&&& km s$^{-1}$ & km s$^{-1}$  & km s$^{-1}$\\
\midrule
232.590, -20.613 & J15302162-2036481  & 6253617213575749376  & -30.20$\pm$0.90  & -14.30$\pm$1.60&-17.80$\pm$1.40   \\
233.817, -25.734 & J15351610-2544030  & 6237729751586459392  & -6.00$\pm$1.30  & -17.90$\pm$2.20&-4.70$\pm$1.90   \\
235.278, -26.941 & J15410679-2656263  & 6234377340635038848  & -5.70$\pm$2.30  & -16.90$\pm$3.90 & -5.70$\pm$3.40  \\
... & ... & ... & ... &... & ... \\
\bottomrule
\end{tabular*}
\begin{tablenotes}%%[341pt]
%\item Source: Example for table source text.
%\item[1] Example for a first table footnote.
%\item[2] Example for a second table footnote.
\end{tablenotes}
\end{table*}
\end{center}

2) Program II: was created using DR2 data \citep{2018A&A...616A...1G} and new spectroscopy of probable members to significantly enhance the USC census within the Sco–Cen association by \cite{2020AJ....160...44L}\footnote{\url{https://vizier.cds.unistra.fr/viz-bin/VizieR?-source=J/AJ/160/44}}, they utilized the latest Gaia DR2 to update the most recent USC census compiled by \cite{2018AJ....156...76L} and identified probable young stars within the Sco–Cen association \citep{1999AJ....117..354D} using a color-magnitude diagram built using Gaia DR2 data. To establish kinematic criteria for membership within that population, they focused on probable members within the USC association’s center concentration. These kinematic criteria, applied to Gaia DR2, were used to revise the list of possible USC members initially chosen by \cite{2018AJ....156...76L}. They aim to finalize the analysis of candidates covering spectral types up to $\sim L0$ ($\sim0.01~M_{\odot}$) for regions included in the United Kingdom Infrared Telescope Infrared Deep Sky Survey (UKIDSS) \citep{2007MNRAS.379.1599L}, which essentially includes most of USC. Based on updated data from DR3 \citep{2023A&A...674A...1G}, identified approximately 209 USC association members with different spectral types, including A- (7 points), B- (17 points), F- (24 points;), G- (19 points), K- (59 points), and M-types (83 points). For kinematic analysis, the analysis focused on the last two spectral types (i.e., K and M types) comprising 142 stars, as the sample sizes for the first four types are relatively small. The data for these stars are presented in Table \ref{tab2}, following the same format as in Table \ref{tab1}.

\begin{center}
\begin{table*}[t]%
\caption{Program II: Spatial velocities for 59 K-type and 83 M-type of USC association \citep{2020AJ....160...44L}}
\label{tab2}
\centering
\begin{tabular*}{500pt}{@{\extracolsep\fill}lcccccD{.}{.}{3}c@{\extracolsep\fill}}
\toprule

($\alpha,~\delta$)$^{o}_{2016}$ & 2MASS  & Gaia DR3  & $U\pm\sigma_{U}$  & $V\pm\sigma_{V}$ & $W\pm\sigma_{W}$ \\
&&& km s$^{-1}$ & km s$^{-1}$  & km s$^{-1}$\\
\bottomrule
 
 &   & K-type; 59 points &   &    \\
\midrule
240.898, -22.766 &	J16033550-2245560 &	6243154501445899264 &		-10.28 $\pm$	0.17 &	-15.52 $\pm$	0.24 &	-9.04 $\pm$	0.12\\
243.327, -22.214 &	J16131858-2212489 &	6242687277719794176 &		-6.37 $\pm$ 0.37 &	-14.77 $\pm$ 0.23&	-8.10 $\pm$ 0.18\\
244.381, -23.060 & J16173138-2303360 &	6050373760491905920 &		-3.92 $\pm$ 0.38 & -14.66 $\pm$ 0.11	&-5.41 $\pm$ 0.15\\

... & ... & ... & ... &... & ... \\
\bottomrule
 &   & M-type; 83 points  &   &    \\
\midrule
236.522, -29.348	& J15460529-2920531	& 6233037070321568128		&-3.44    $\pm$	3.14	&-16.61 $\pm$	0.99	&-8.75 $\pm$	1.18\\

239.579, -20.907	& J15581906-2054238	& 6246706714275425024 &-5.88 $\pm$ 1.05 & -16.03 $\pm$ 0.21 & -6.58 $\pm$ 0.48\\

241.381, -19.762 & J16053153-1945435 & 6247208950571840640 & 3.83 $\pm$ 1.64 & -14.11 $\pm$ 0.27	& -6.36 $\pm$ 0.73\\

... & ... & ... & ... &... & ... \\
\bottomrule
\end{tabular*}
\begin{tablenotes}%%[341pt]
%\item Source: Example for table source text.
%\item[1] Example for a first table footnote.
%\item[2] Example for a second table footnote.
\end{tablenotes}
\end{table*}
\end{center}

3) Program III: \cite{2022A&A...667A.163M}\footnote{\url{https://vizier.cds.unistra.fr/viz-bin/VizieR?-source=J/A+A/667/A163}} identified seven different groups in the USC and Oph star forming regions. Due to the proximity of USC to the Oph star-forming clouds, no ongoing star formation is indicated. The molecular clouds in this region have already dispersed, although members of varying masses are still observable \citep{2018MNRAS.477.5191R}. 

Two groups ($\rho$ Oph and $\alpha$ Sco) are located in Oph, four groups ($\gamma$ Sco, $\beta$ Sco, $\sigma$ Sco, and $\delta$ Sco) are in USC and share a common origin, and the seventh group ($\pi$ Sco) represents a nearby young population. These groups exhibit an age gradient, indicating that star creation in the region has been a sequential process throughout the previous 5 Myr, ranging from the youngest Oph group to the oldest Sco group ($\lesssim$ 5 Myr). Their trace-back analysis revealed the Oph and USC groups share a common parent.

Using the Hipparcos or DR3 catalogs, the USC and Oph member stars formed the initial sample, as selected by researchers \citep{2022NatAs...6...89M, 2022A&A...667A.163M}. The present study identifies approximately 670 probable DR3 members from these associations, divided into seven groups: i) $\alpha$ Sco (135 stars), ii) $\beta$ Sco (61 stars), iii) $\delta$ Sco (131 stars), iv) $\sigma$ Sco (104 stars), v) $\mu$ Sco (57 stars), vi) $\pi$ Sco (88 stars), and vii) $\rho$ Oph (94 stars). The data for these members are presented in Table \ref{tab3}, with the following format: the first column represents the equatorial coordinates ($\alpha,~\delta$)\textsubscript{2016}, the second column represents Gaia DR3 identifier number, and the last three columns display the spatial velocities with their uncertainty errors relative to the Galactic coordinates.

Figure \ref{fig1} illustrates the distribution of radial velocities ($V_r$; km \textsuperscript{-1}) as a function of Galactic longitude ($l^o$) for Programs I, II, and III.

\begin{center}
\begin{table*}[t]%
\caption{Program III: 670 members of USC and Oph associations for seven identified groups \citep{2022A&A...667A.163M}\label{tab3}}
\centering
\begin{tabular*}{500pt}{@{\extracolsep\fill}lcccccD{.}{.}{3}c@{\extracolsep\fill}}
\toprule

%&\multicolumn{2}{@{}c@{}}{\textbf{Spanned heading\tnote{1}}} & \multicolumn{2}{@{}c@{}}{\textbf{Spanned heading\tnote{2}}} \\\cmidrule{2-3}\cmidrule{4-5}
$(\alpha,~\delta)^{o}_{2016}$   & Gaia DR3  & $U\pm\sigma_{U}$  & $V\pm\sigma_{V}$ & $W\pm\sigma_{W}$ \\
&& km s$^{-1}$ & km s$^{-1}$  & km s$^{-1}$\\
\bottomrule
 
 &   & $\alpha$ Sco; 135 stars &   &    \\
\midrule
244.872, -21.404 &	6244295760151546624&	-2.68$\pm$	0.45&	-16.71$\pm$	0.07&	-5.86 $\pm$	0.17\\
243.633, -28.123	&6042179203410212864&	-3.43 $\pm$	0.10&	-16.29$\pm$	0.07&	-5.52 $\pm$	0.05\\
246.460, -28.344	&6044217320011144832&	-2.86$\pm$	0.10&	-16.12$\pm$	0.13&	-5.472$\pm$	0.07\\

... & ... & ... & ... & ... \\
\bottomrule

 &   & $\beta$ Sco; 61 stars &   &    \\
\midrule

241.610,	-23.103		&6242336396077230720&	-2.68 $\pm$	0.21 &	-16.98$\pm$	0.20 &	-6.88$\pm$	0.12\\
241.705,	-22.277		&6242541558069369216&	-0.83$\pm$	0.39 &	-16.76$\pm$	0.39 &	-7.29$\pm$	0.26\\
243.556,	-22.750 		&6242595262340274688&	0.19$\pm$0.15	&-17.01$\pm$	0.12 &	-7.00$\pm$	0.08\\

... & ... & ... & ... & ... \\
\bottomrule

&   & $\delta$ Sco; 131 stars &   &    \\
\midrule
240.845,	-25.755	&6043652720795418496	&	-5.89$\pm$	0.09 &	-16.79 $\pm$	0.06&	-7.44$\pm$	0.04\\
244.227,	-25.431	&6048832623143380480	&	-6.21$\pm$	0.25&	-16.08$\pm$	0.24&	-10.07$\pm$	0.16\\
243.725,	-25.075	&6049525830867571072	&	-7.00	$\pm$0.12&	-15.89$\pm$	0.11&	-9.51$\pm$	0.07\\
... & ... & ... & ... & ... \\
\bottomrule

&   & $\sigma$ Sco; 104 stars &   &    \\
\midrule
238.341,	-21.971 	&6240546906538196352	&	-4.94	$\pm$	0.21 &	-16.32	$\pm$	0.09	&	-7.90	$\pm$	0.10\\
244.105,	-24.989	&6049614307192716288	&	-7.74	$\pm$	0.56&	-16.96	$\pm$	0.12	&	-6.35	$\pm$	0.19\\
245.846,	-29.026	&6038188148001109248	&	-6.04	$\pm$	0.23&	-14.73	$\pm$	0.19	&	-6.03	$\pm$	0.10\\

... & ... & ... & ... & ... \\
\bottomrule

&   & $\mu$ Sco; 57 stars &   &    \\
\midrule
242.261,	-18.996&	6248783275121192704&		-6.79	$\pm$	0.88&	-14.16	$\pm$	0.10&	-9.92	$\pm$	0.39\\
243.091,	-19.579	&6245692620954228864&		-9.04	$\pm$	0.52&	-14.81	$\pm$	0.08&	-9.66	$\pm$	0.22\\
243.095,	-21.455	&6243054686400695040&		-5.10	$\pm$	0.25&	-15.79	$\pm$	0.22&	-9.43	$\pm$	0.16\\

... & ... & ... & ... & ... \\
\bottomrule

&   & $\pi$ Sco; 88 stars &   &    \\
\midrule
236.847,	-25.868&	6234897925031130624&		-8.90$\pm$	0.49&	-17.53$\pm$	0.15&	-5.63$\pm$	0.21\\
242.541,	-24.583&	6049847575458556800&		-4.72$\pm$	0.89&	-18.96$\pm$	0.17&	-5.53$\pm$	0.32\\
243.025,	-23.246&	6242180368499938816&		-8.39	$\pm$0.62&	-16.75$\pm$	0.10&	-4.54$\pm$	0.23\\

... & ... & ... & ... & ... \\
\bottomrule

&   & $\rho$ Oph; 94 stars &   &    \\
\midrule
245.209, -22.594 & 6050496459111584384 & -8.65 $\pm$ 0.33 &	-13.88 $\pm$ 0.06 & -6.38 $\pm$ 0.12\\
246.386, -26.194 & 6045872806564888192 & -5.12 $\pm$ 0.20 &	-14.68 $\pm$ 0.10 & -9.15 $\pm$ 0.08\\
247.266, -24.862 & 6046072986395944064 & -7.59 $\pm$ 0.20 &	-15.37 $\pm$ 0.18 & -8.75 $\pm$ 0.12\\

... & ... & ... & ... & ... \\
\bottomrule

\end{tabular*}
\begin{tablenotes}%%[341pt]
%\item Source: Example for table source text.
%\item[1] Example for a first table footnote.
%\item[2] Example for a second table footnote.
\end{tablenotes}
\end{table*}
\end{center}

\begin{figure}
\label{fig1}
\centerline{\includegraphics[width=0.5\textwidth,clip=]{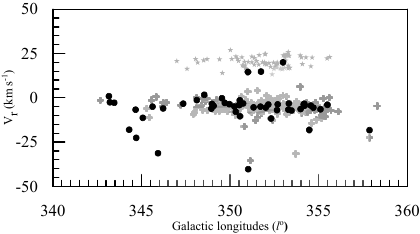}}
\caption{Distribution of radial velocities ($V_r$;
km s\textsuperscript{-1}) of USC and Oph member association stars versus their Galactic longitudes ($l^o$) for the three Programs. Data points are represented as black closed circles (Program I), middle gray closed pluses (Program II), and faint gray closed stars (Program III).}

\label{fig1}
\end{figure}

\section{Computational methods}
\label{sec3}
\subsection{Inner kinematics and VEPs}
Let $d$(pc) be the distance from the Sun to the star, and, because the observation data for proper motions are always given in the equatorial system, let us resolve positions and velocities into equatorial rectangular components \citep{Smart68},

\begin{equation}
\label{Eq._xyz}
\begin{split}
x~&=~d~\cos\delta~\cos\alpha,\\
y~&=~d~\cos{\delta}~\sin{\alpha},\\
z~&=~d~\sin{\delta}.
\end{split}
\end{equation}

Differentiation Eqs. (\ref{Eq._xyz}) with respect to time, we have the equatorial components of the stellar velocity with respect to the Sun.

\begin{equation}
\label{eq:Vxyz}
\begin{split}
V_x & = \Dot{d}~\cos\delta \cos\alpha~-~\Dot{\alpha}d \cos\delta \sin\alpha~-~\Dot{\delta}d \sin\delta \cos\alpha , \\ %\\%vspace{1cm}\\
V_y & = \Dot{d}~\cos\delta \sin\alpha~+~\Dot{\alpha}d \cos\delta \cos\alpha~-~\Dot{\delta}d \sin\delta \sin\alpha, \\ %\\%vspace{1cm}\\
V_z & = \Dot{d}~\sin\delta ~+~\Dot{\delta}d \cos{\delta}. \\ %\\%vspace{1cm}\\
\end{split}
\end{equation}

The quantity $\Dot{d}$ stands for the radial velocity ($V_r$; km s$^{-1}$) and $\Dot{\alpha}d\cos{\delta}$ and $\Dot{\delta}d$ are traverse velocities (i.e., $\Dot{\alpha}d\cos{\delta}=V_{\alpha}=4.74d\mu_{\alpha}\cos{\delta}$ and $\Dot{\delta}d=V_{\delta}=4.74d\mu_{\delta}$; km s$^{-1}$) and may be written as 

\begin{equation}
\begin{bmatrix}
V_x\\
V_y\\
V_z
\end{bmatrix}
=
\begin{bmatrix}
V_r\\
V_{\alpha}\\
V_{\delta}
\end{bmatrix}
.
\begin{bmatrix}
\cos{\alpha}\cos{\delta} & -\sin{\alpha} & -\cos{\alpha}\sin{\delta}\\
\sin{\alpha}\cos{\delta} & \cos{\alpha} & -\sin{\alpha}\sin{\delta}\\
\sin{\delta} & 0 & \cos{\delta}
\end{bmatrix}
\end{equation}

To facilitate the analysis, we must solve Eqs. (\ref{eq:Vxyz}) for $V_r$, $V_{\alpha}$, and $V_{\delta}$, so that we can write equations that involve only radial velocities or only proper motions.

\begin{equation}
\label{eq:Vxyz2}
\begin{split}
V_x~\cos\delta \cos\alpha~+~V_y~\cos\delta \sin\alpha~+~V_z~\sin\delta & = V_r, \\ %\\%vspace{1cm}\\
-V_x~\sin\delta \cos\alpha~-~V_y~\sin\delta \sin\alpha~+~V_z~\cos\delta & = 4.74d\mu_{\delta}, \\ %\\%vspace{1cm}\\
-V_x~\sin\alpha~+~V_y~\cos\alpha & = 4.74d\mu_{\alpha}\cos{\delta}. \\ %\\%vspace{1cm}\\
\end{split}
\end{equation}

The transformation of equatorial velocities in Eqs. (\ref{eq:Vxyz}) to Galactic velocities ($U,~V,~W$; km s$^{-1}$), being determined by the definition of the equatorial positions of the North Galactic Pole and zero Galactic longitude, therefore, according to \cite{1987AJ.....93..864J}.

\begin{equation}
\label{eq:B}
\begin{bmatrix}
U\\
V\\
W
\end{bmatrix}
=B.
\begin{bmatrix}
V_r\\
V_{\alpha}\\
V_{\delta}
\end{bmatrix}
\end{equation}

where $B=T.A$,\\

$T=$
$\begin{bmatrix}
-0.06699 & -0.87276 & -0.48354\\
+0.49273 & -0.45035 & +0.74458\\
-0.86760 & -0.18837 & +0.46020
\end{bmatrix}$\\

and\\ 

$A\equiv$
$\begin{bmatrix}
+\cos{\alpha} \cos{\delta} & -\sin{\alpha} & -\cos{\alpha} \sin{\delta}\\
+\sin{\alpha} \cos{\delta} & +\cos{\alpha} & -\sin{\alpha} \sin{\delta}\\
+\sin{\delta} & 0 & +\cos{\delta}
\end{bmatrix}$

The solution is easy to write because the matrix of coefficients in Eqs. ($\ref{eq:B}$) is orthogonal, so that its inverse equals its transpose. We find that
\begin{equation}
\label{eq:UVW1}
\begin{split}
U &=-0.06699~V_{x}-0.87276~V_{y}-0.48354~V_{z},\\ %\\
V &=+0.49273~V_{x}-0.45035~V_{y}+0.74458~V_{z},\\ %\\
W &=-0.86760~V_{x}-0.18837~V_{y}+0.46020~V_{z}.%\\
\end{split}
\end{equation}

On the other hand, the derivations are well defined considering Eqs. (\ref{eq:Vxyz}) by using an equatorial-Galactic transformation matrix based on the SPECFIND v2.0 catalog of radio continuum spectra; see Eq. (14) in \cite{Liu11} (printed to 10 decimals, 1 mas accuracy):
\begin{equation}
\label{eq:UVW2}
\begin{split}
U &=-0.0518807421~V_{x}-0.8722226427~V_{y} \\
&-0.4863497200~V_{z},\\ %\\
V &=+0.4846922369~V_{x}-0.4477920852~V_{y} \\
&+0.7513692061~V_{z},\\ %\\
W &=-0.8731447899~V_{x}-0.1967483417~V_{y} \\
&+0.4459913295~V_{z},%\\
\end{split}
\end{equation}

To specify VEPs as outlined in the literature \citep{elsanhoury24,elsanhouryAN}, considering estimation of the velocities dispersion ($\sigma_1,~\sigma_2$ and $\sigma_3$; km s$^{-1}$) using the following expressions:

\begin{equation}
\begin{array}{l} {\sigma_1}=\sqrt{2\rho^{\frac{1}{3}}\cos\frac{\phi}{3}-\frac{k_{1}}{3};}\\
{\sigma_2}=\sqrt{-\rho^{\frac{1}{3}}\left\{\cos\frac{\phi}{3}+\sqrt{3}\sin\frac{\phi}{3}\right\}-\frac{k_{1}}{3};}\\{\sigma_3}
=\sqrt{-\rho^{\frac{1}{3}}\left\{\cos\frac{\phi }{3}-\sqrt{3}\sin\frac{\phi}{3}\right\}-\frac{k_{1}}{3}}.{\rm \; \; \;
\; \; \; \; \; \; \; \; \; \; \; \; \; \; \; \; \; \; \; \; \; } \end{array}
\end{equation}

The parameters $q$ and $r$ are given by the equations:

\begin{equation}
q=\frac{1}{3} k_{2} -\frac{1}{9} k_{1}^{2} {\rm \; \; \; \; \; \; ;\; \; }r=\frac{1}{6} \left(k_{1} k_{2} -3k_{3} \right)-\frac{1}{27} k_{1}^{3}
\end{equation}

$\rho$ and $\phi$ are calculated as:

\begin{equation}
\rho=\sqrt{-q^{3} },
\end{equation}
\begin{equation}
x=\rho ^{2}-r^{2},
\end{equation}
\begin{equation}
\phi
=\tan ^{-1} \left(\frac{\sqrt{x} }{r} \right).
\end{equation}

The coefficients $k_1$, $k_2$, and $k_3$  are determined as:
\begin{equation}
\begin{array}{l} {k_{1}=-\left(\mu _{11} +\mu _{22} +\mu _{33} \right),} \\ {k_{2}=\mu _{11} \mu _{22} +\mu _{11} \mu _{33} +\mu _{22} \mu _{33} -\left(\mu _{12}^{2} +\mu _{13}^{2} +\mu _{23}^{2} \right),} \\ {k_{3}=\mu _{12}^{2} \mu _{33} +\mu _{13}^{2} \mu _{22} +\mu _{23}^{2} \mu _{11} -\mu _{11} \mu _{22} \mu _{33} -2\mu _{12} \mu _{13} \mu _{23} .} \end{array}
\end{equation}

and the matrix elements ($\mu_{ij}$) are 
\begin{equation}
 \begin{array}{l} {\mu _{11}=\frac{1}{N} \sum _{i=1}^{N}U_{i}^{2}  -\left(\overline{U}\right)^{2} ;{\rm \; \; \; \; \; }\mu _{12}=\frac{1}{N} \sum _{i=1}^{N}U_{i} V_{i}  -\overline{U}\; \overline{V};} \\ {\mu _{13}=\frac{1}{N} \sum _{i=1}^{N}U_{i} W_{i}  -\overline{U}\; \overline{W};{\rm \; \; }\mu _{22}=\frac{1}{N} \sum _{i=1}^{N}V_{i}^{2}  -\left(\overline{V}\right)^{2} ;} \\ {\mu _{23}=\frac{1}{N} \sum _{i=1}^{N}V_{i} W_{i}  -\overline{V}\; \overline{W};{\rm \; \; \; }\mu _{33}=\frac{1}{N} \sum _{i=1}^{N}W_{i}^{2}  -\left(\overline{W}\right)^{2} .} \end{array}
 \end{equation}

Accordingly, Figure $\ref{fig2}$ presents the distribution of the Galactic spatial velocities components ($U,~V,~W$) for our retrieved three Programs with their figures keys.

\begin{figure}
\label{fig2}
\centerline{\includegraphics[width=0.5\textwidth,clip=]{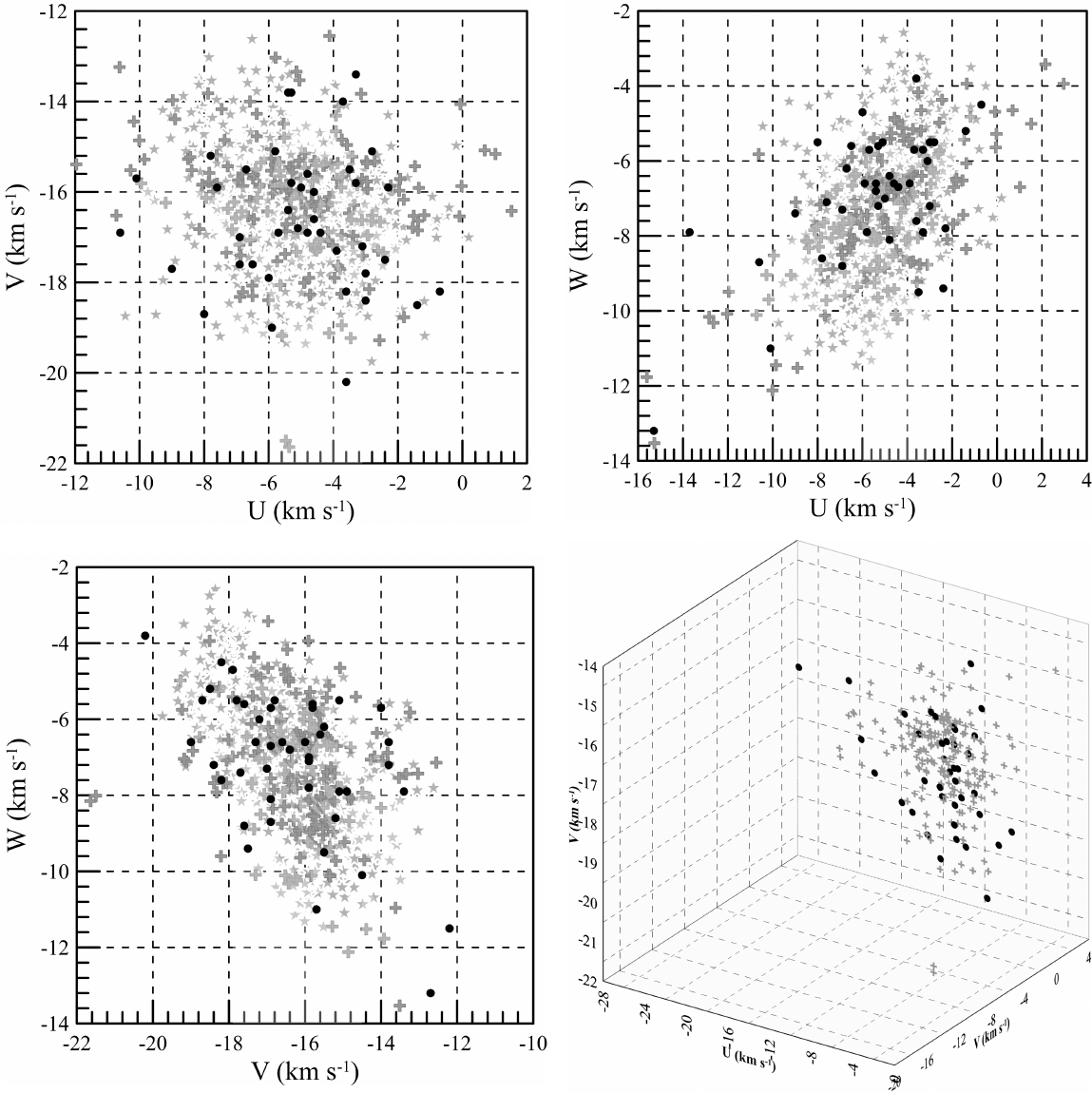}}
\caption{Distribution of spatial velocity components: ($V~vs.~U$), ($W~vs.~U$), and ($W~vs.~V$) along the Galactic coordinates for three Programs. Data points are represented as black closed circles (Program I), middle gray closed pluses (Program II), and faint gray closed stars (Program III). The figure also displays their 3D array representation.}
\label{fig2}
\end{figure}

\subsection{CP}

Nearby star associations that share a common space motion exhibit proper motions that seem to "converge" towards one specific sky point and share similarities within the Galaxy as they are gravitationally bound. Therefore, the parameters such as age, distance, chemical composition, and kinematics (internal motion or trajectory) are correlated. The converge can be understood as accumulations of stars moving with a common velocity vector, their parallel motions on the celestial sphere will direct them toward a coherent point, known as the vertex, apex, or convergent point (CP) of the open clusters (OCs) and/or associations. Computing the apex position assumes that the moving group is neither expanding, contracting, nor rotating, and that its motion relative to the field is significant enough to enable accurate membership discrimination. \cite{Vereshchagin14}, \cite{Elsanhoury18}, \cite{Bisht20}, \cite{Elsanhoury21}, and \cite{Maurya21} used the method of individual star apex (AD-diagram) as described by \cite{Chupina01, Chupina06}. This method utilizes average space velocity vectors, as expressed in Eqs. (\ref{eq:Vxyz}), to determine the equatorial coordinates of the CP ($A_o,~D_o$) in the following forms:

\begin{equation}
\label{eq:ad}
\begin{split}
A_{o} &= \tan^{-1} \Bigg(\frac{\overline{ V_y}}{\overline{V_x}}\Bigg) \\
%\end{equation}
%\begin{equation}
D_{o} &= \tan^{-1} \Bigg(\frac{\overline{V_z}}{\sqrt {\overline{V_x^2} + \overline{V_{y}^{2}}}}\Bigg)
\end{split}
\end{equation}

The cross mark in Figure \ref{fig3} shows the apex position of Program III as listed in Table \ref{tab4}.

\begin{figure}
\label{fig3}
\centerline{\includegraphics[width=0.5\textwidth,clip=]{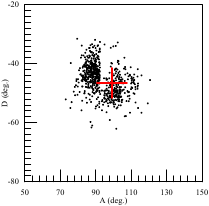}}
\caption{AD-diagram of Program III associations, showing 670 points with the cross point ($A_o,~D_o$) indicated.}
\label{fig3}
\end{figure}

\subsection{Solar motion elements}
Consider a group with mean spatial velocities ($\overline{U}$, $\overline{V}$, $\overline{W}$) along the Galactic coordinates, based on observational data. The components of the Sun’s velocities ($U_\odot$, $V_\odot$ and $W_\odot$) are expressed as follows: $U_\odot=-\overline{U}=-S_\odot\cos{l_A} \cos{b_A}$, $V_\odot=-\overline{V}=S_\odot\cos{l_A} \sin{b_A}$, and $W_\odot=-\overline{W}=S_\odot\sin{b_A}$. As a result, we determine the Solar elements, i.e.,

\begin{equation}
S_{\odot}=\sqrt{\overline{U}^2+\overline{V}^2+\overline{W}^2},
\end{equation}
\begin{equation}
l_{A}=tan^{-1}\Bigg(\frac{-\overline{V}}{\overline{U}}\Bigg),
\end{equation}
\begin{equation}
b_{A}=sin^{-1}\Bigg(\frac{-\overline{W}}{S_{\odot}}\Bigg).
\end{equation}

The Sun's velocities were relative to these groups and the same axes are defined as ($X_{\odot}^{\bullet}=-\overline{V_x}$), ($Y_{\odot}^{\bullet}=-\overline{V_y}$), and ($Z_{\odot}^{\bullet}=-\overline{V_z}$), taking into account the position along the $x,y,$ and $z$ - axes in the coordinate system centered around the Sun (i.e., equatorial coordinates). Consequently, we are obtained with Solar components that have radial velocities.

\begin{equation}
S_{\odot}=\sqrt{\left(X_{\odot}^{\bullet}\right)^2+\left(Y_{\odot}^{\bullet}\right)^2+\left(Z_{\odot}^{\bullet}\right)^2},
\end{equation}
\begin{equation}
\alpha_{A}=tan^{-1}\Bigg(\frac{Y_{\odot}^{\bullet}}{X_{\odot}^{\bullet}}\Bigg),
\end{equation}
\begin{equation}
\delta_{A}=tan^{-1}\Bigg(\frac{Z_{\odot}^{\bullet}}{\sqrt{\left(X_{\odot}^{\bullet}\right)^2+\left(Y_{\odot}^{\bullet}\right)^2}}\Bigg).
\end{equation}

\begin{sidewaystable}%[h]
\tiny
\caption{The kinematics, VEPs, Solar elements, and the convergent points retrieved for Programs I, II, and III.}%
\begin{tabular*}{\textheight}{@{\extracolsep\fill}lcccccccccD{.}{.}{4}D{.}{.}{4}D{.}{.}{4}D{.}{.}{4}D{.}{.}{4}D{.}{.}{4}@{\extracolsep\fill}}%
\toprule
\label{tab4}
 Program & $N$ & ($\overline{V_x},~\overline{V_y},~\overline{V_z}$; km s$^{-1}$) & ($\overline{U},~\overline{V},~\overline{W}$; km s$^{-1}$)  & ($\sigma_1,~\sigma_2,~\sigma_3$; km s$^{-1}$) & ($S_\odot$; km s$^{-1}$) &($l_{A},~b_{A}$)$^{o}$&($\alpha_{A},~\delta_{A}$)$^{o}$&($A_{o},~D_{o}$)$^{o}$\\
\midrule

I	&	51	&	-1.10$\pm$0.03, 14.96$\pm$3.87, -11.90$\pm$3.45	&	-7.21$\pm$2.69, -16.17$\pm$4.02, -7.29$\pm$2.70	&	19.60$\pm$4.43, 10.41$\pm$3.23, 1.63$\pm$0.78	&	19.14$\pm$4.37	&	-65.97,22.38 	&	-85.80, 38.43		&	94.21$\pm$0.11, -38.43$\pm$0.02\\
II; K-type	&	59	&	-1.18$\pm$0.03, 12.72$\pm$3.57, -12.77$\pm$3.57	&	-4.82$\pm$0.46, -15.85$\pm$3.98, -7.17$\pm$2.68	&	17.95$\pm$4.23, 2.61$\pm$0.62, 1.14$\pm$0.94	&	18.06$\pm$4.25	&	-73.10, 23.40	&	-84.70, 45.00		&	95.30$\pm$0.10, -50.00$\pm$0.01\\
II; M-type	&	83	&	-0.81$\pm$0.03, 13.66$\pm$3.70, -12.76$\pm$3.57	&	-5.67$\pm$0.42, -16.10$\pm$4.01, -7.67$\pm$2.77	&	18.64$\pm$4.32, 2.56$\pm$0.63, 1.19$\pm$0.09	&	18.71$\pm$4.33	&	-70.61, 24.21	&	-86.61, 43.00		&	93.40$\pm$0.10, -43.00$\pm$0.02\\
III; $\alpha$ Sco	&	135	&	-2.78$\pm$0.60, 11.68$\pm$3.42, -13.38$\pm$3.66	&	-3.54$\pm$0.53, -16.63$\pm$4.08, -5.85$\pm$0.41	&	1.00$\pm$0.00, 0.70$\pm$0.01, 0.60$\pm$0.01,	&	18.00$\pm$4.24	&	-78.00, 18.98	&	-76.62, 48.10		&	103.39$\pm$0.09, -48.10$\pm$0.01\\
III; $\beta$ Sco	&	61	&	-1.54$\pm$0.03, 11.36$\pm$3.37, -13.71$\pm$3.70	&	-3.17$\pm$0.56, -16.13$\pm$4.02, -7.00$\pm$2.65	&	1.10$\pm$0.01, 0.50$\pm$,0.00 0.70$\pm$0.01	&	17.87$\pm$4.23	&	-78.90, 23.07	&	-82.28, 50.10		&	97.72$\pm$0.10, -50.10$\pm$0.01\\
III; $\delta$ Sco	&	131	&	-0.80$\pm$0.03, 14.25$\pm$3.77, -12.58$\pm$3.55	&	-6.26$\pm$2.50, -16.22$\pm$4.03, -7.72$\pm$2.78	&	0.80$\pm$0.01, 0.60$\pm$0.01, 0.75$\pm$0.01	&	19.02$\pm$4.36	&	-68.88, 23.94	&	-86.79, 41.40		&	93.21$\pm$0.10,-41.40$\pm$0.02\\
III; $\sigma$ Sco	&	104	&	-1.12$\pm$0.03, 13.82$\pm$3.72, -12.95$\pm$3.60	&	-5.69$\pm$2.38, -16.46$\pm$4.06, -7.52$\pm$2.74	&	2.20$\pm$0.01, 1.50$\pm$0.01, 1.30$\pm$0.01	&	18.97$\pm$4.36	&	-70.92, 23.35	&	-85.37, 43.05		&	94.64$\pm$0.10, -43.05$\pm$0.02\\
III; $\mu$ Sco	&	57	&	0.44$\pm$0.01, 13.71$\pm$3.70, -12.53$\pm$3.54	&	-5.88$\pm$2.42, -15.34$\pm$3.92, -8.67$\pm$2.94	&	1.00$\pm$0.01, 0.50$\pm$0.01, 1.10$\pm$0.01	&	18.58$\pm$4.31	&	-69.01, 27.82	&	88.16, 42.41		&	88.17$\pm$0.11, -42.41$\pm$0.02\\
III; $\pi$ Sco	&	88	&	-4.42$\pm$0.48, 13.25$\pm$3.64, -13.32$\pm$3.65	&	-4.85$\pm$0.45, -18.08$\pm$4.25, -4.69$\pm$0.46	&	1.80$\pm$0.01, 0.90$\pm$0.01, 1.00$\pm$0.01	&	19.30$\pm$4.40	&	-75.00, 14.08	&	-71.55, 43.64		&	108.45$\pm$0.01, -43.64$\pm$0.02\\
III; $\rho$ Pho	&	94	&	1.01$\pm$0.01, 13.74$\pm$3.71, -12.51$\pm$3.54	&	-5.96$\pm$0.41, -15.06$\pm$3.88, -9.16$\pm$3.03	&	1.60$\pm$0.01, 0.90$\pm$0.01, 1.20$\pm$0.02	&	18.61$\pm$4.32	&	-68.43, 29.50	&	85.80, 42.24		&	85.80$\pm$0.11, -42.24$\pm$0.01\\

\bottomrule
\end{tabular*}
\begin{tablenotes}%%[\textheight]

\end{tablenotes}
\end{sidewaystable}

\subsection{Rotational A $\&$ B constants}
Many kinematic and photometric parameters can be extracted from OCs and/or stellar associations with great accuracy. These parameters are essential for studying the Galaxy and its subsystems. Nearby stellar associations serve as valuable tools for investigating the characteristics of the thick and thin disks of the Galaxy, including their chemical and dynamical history, spiral structure, star formation, and distance scale determination.

Based primarily on the analysis of spatial positions, the morphology of the stellar components of the Galaxy has been refined. By examining the velocities of stars, particularly those near the Sun with well-determined coordinates and velocities, one can establish a relationship between the morphological parameters of these stellar groups and the kinematic properties of the stellar population of the Galaxy.

Traditionally, the study of the stellar structure of the Galaxy has relied on the position of stars in the three spatial coordinates $x$, $y$, and $z$ \citep{2023ARep...67.1418T}. The components of the stellar spatial velocities ($U$, $V$, and $W$) can be determined using radial velocity spectral data, long-term observations of stellar motion in the celestial sphere with known parallaxes, and consideration of the Sun’s motion within the Galaxy. DR3 catalog \citep{2023A&A...674A...1G} which provides data for approximately two billion stars in the Galaxy is particularly effective in predicting these values.

For nearly a century, the kinematics of nearby stars have been utilized to investigate the characteristics of the Galactic disk. Oort demonstrated that the constants of the stream velocity field equation could describe the local frequency and the circular velocity gradient \citep{OortA,oortb}. He used radial velocities and proper motions to determine the Oort rotational constants ($A~\&~B$; km s$^{-1}$ kpc$^{-1}$), arriving at values of $A\approx$ 19 km s$^{-1}$ kpc$^{-1}$ and $B\approx$ -24 km s$^{-1}$ kpc$^{-1}$. This study refuted the notion of solid-body rotation for the Galaxy, revealing a nearly flat rotation curve. Computation of several parameters is achieved, including ratios ($\sigma_2⁄\sigma_1$) of Programs I, II, III, and Oort's constants.\\
i.e.

\begin{equation}
\label{sigma12}
\Bigg(\frac{\sigma_{2}}{\sigma_{1}}\Bigg)^2=\frac{-B}{A-B},
\end{equation}

or

\begin{equation}
~~~\frac{-B}{A}=\frac{1}{(\sigma_{1}/\sigma_{2})^2-1}.
\end{equation}

\section{Kinematic and rotation results}
\label{sec4}

Following the above computational scheme, a Mathematica routine was created and developed to compute the kinematics and VEPs of Programs I, II, and III in addition to Oort's constants.

Table \ref{tab4} lists our original numerical results for three Programs, including various kinematic parameters of space velocities ($\overline{V_x},~\overline{V_y},~\overline{V_z}$) and their corresponding spatial velocities $\overline{U}=\frac{1}{N}\sum_{i=1}^{N}U_{i}$, $ \overline{V}=\frac{1}{N}\sum_{i=1}^{N}V_{i}$, and $\overline{W}=\frac{1}{N}\sum_{i=1}^{N}W_{i}$ along equatorial and Galactic coordinates, respectively, velocities dispersions ($\sigma_1,~\sigma_2,~\sigma_3$; km s$^{-1}$), Solar elements ($S_\odot$, $l^{o}_A$, $b^{o}_A$, $\alpha^{o}_A$, $\delta^{o}_A$), CP coordinates ($A_o, D_o$)$^{o}$, and Oort's constants.

Mean spatial velocities ($\overline{U},~\overline{V}$, and $\overline{W}$) were calculated, and their combined value is as follows: $V_{space}=(\overline{U^2}+\overline{V^2}+\overline{W^2})^{1/2}$ (km s$^{-1}$) for USC Program I (-7.21 $\pm$ 2.69, -16.17 $\pm$ 4.02, -7.29 $\pm$ 2.70, and 19.15 $\pm$ 4.38 km s$^{-1}$), Program II (-5.25 $\pm$ 2.29, -15.98 $\pm$ 4.00, -7.42 $\pm$ 2.72, and 18.38 $\pm$ 4.29 km s$^{-1}$), and Program III are all associated with USC and Oph (i.e., -5.05 $\pm$ 2.25, -16.27 $\pm$ 4.03, -7.23 $\pm$ 2.69, and 18.51 $\pm$ 4.30 km s$^{-1}$). These values coincide with those calculated by \cite{2018MNRAS.477L..50G} wherein ($\overline{U}=$ -5.00 $\pm$ 0.10, $\overline{V}=$ -16.60 $\pm$ 0.10, $\overline{W}=$ -6.80 $\pm$ 0.10, and 18.80 $\pm$ 0.10 km s$^{-1}$), according to \cite{2022A&A...667A.163M} (-4.97, -16.27, -7.21) and their $V_{space}$ is 18.48 km s$^{-1}$, for 209 stars that were accepted as members of USC \citep{2020AJ....160...44L}, the mean spatial velocities are ($\overline{U},~\overline{V},~\overline{W}$) = (-5.10 $\pm$ 4.10, -16.00 $\pm$1.40, -7.20 $\pm$2.10), and $\overline{U}=-6.16^{+0.14}_{-0.13},~\overline{V}=-16.89^{+0.08}_{-0.10},~\overline{W}=-7.05^{+0.09}_{-0.08}$ km s$^{-1}$ within DR1 for a smaller sample of members by \cite{2018MNRAS.476..381W}.

Mean velocity dispersion ($\sigma_j$; $\forall{j=1,2,3}$; km s$^{-1}$) where ($\sigma_1>\sigma_2>\sigma_3$). Therefore, $\overline{\sigma_1}$, $\overline{\sigma_2}$, and $\overline{\sigma_3}$ for the three programs are as follows: Program I (19.60 $\pm$ 4.43, 10.41 $\pm$ 3.23, 1.63 $\pm$ 0.78), Program II (18.30 $\pm$ 4.23, 2.60 $\pm$ 0.62, 1.17 $\pm$ 0.93), and Program III (1.36 $\pm$ 0.02, 0.80 $\pm$ 0.01, 0.96 $\pm$ 0.01), $\sigma$'s are around $1.63^{+0.20}_{-0.20}$, $1.14^{+0.13}_{-0.14}$, and $2.51^{+0.11}_{-0.09}$ according to data achieved through DR1 \citep{2018MNRAS.476..381W}. 

The absolute value of the Sun’s velocity is denoted as ($S_\odot$; km s$^{-1}$) considering the retrieved associations under study, where the Solar apex’s Galactic longitude and latitude relative to Galactic center is ($l_A,~b_A$) and in the equatorial coordinates ($\alpha_A,~\delta_A$). The position of the apex ($l_A, b_A$) in the Galactic coordinates can be calculated using basic spherical trigonometry formulas via ($\alpha_{A}$, $\delta_{A}$). Numerically, mean absolute value of the Sun's velocity $\overline{S_\odot}$ relative to our three program associations are 19.14 $\pm$ 4.37 (Program I), 18.39 $\pm$ 4.29 (Program II), and 18.62 $\pm$ 4.32 (Program III). Columns seven and eight of Table \ref{tab4} present estimated ($l_A$, $b_A$) and ($\alpha_A$,~$\delta_A$), respectively.

The last column of Table \ref{tab4} indicate the CP coordinates ($A_o,~D_o$)$^{o}$; 94$^o$.21$\pm$ 0$^o$.11, -38$^o$.43 $\pm$ 0$^o$.02 (Program I), 94$^o$.35 $\pm$ 0$^o$.11, -46$^o$.50 $\pm$ 0$^o$.15 (Program II), and 95$^o$.91 $\pm$ 0$^o$.10, -44$^o$.42 $\pm$ 0$^o$.15 (Program III), all of which are in line with the findings of earlier research by \cite{1999MNRAS.306..381D} into which ($A_o$ = 95$^o$.20 $\pm$ 2$^o$.20, $D_o$ = -42$^o$.80 $\pm$ 2$^{o}$.50), $A_o$ = 110$^o$.97 $\pm$ 2$^o$.20 and $D_o$ = -20$^o$.25 $\pm$ 2$^o$.50 by \cite{2018MNRAS.477L..50G}, and utilizing Jones method \citep{1971MNRAS.152..231J} \cite{2018MNRAS.476..381W} estimated CPs for USC associations as $A_o$ = 116$^o$.22$^{+10.70}_{-9.46}$ and $D_o$ = -55$^o$.29$^{+5.37}_{-4.56}$.

The absolute value of $A$ and $B$ (i.e., |$A-B$|) is defined as the angular rotation rate and therefore the differential Galactic rotation rate with radius $R_o=8.20\pm0.10$ pc \citep{Bland19} was provided by \cite{OortA,oortb} at 247.50 km s$^{-1}$ and mathematically is $V_o=\omega_{o}R_{o}=|A-B|R_{o}$. \cite{1981gask.book.....M} states that $V_{o}$ yields in the range of 200-300 km s$^{-1}$ based on radial velocities of globular clusters or spheroidal component stars in our Galaxy or external galaxies in the Local Group. Recently, $V_{o}$ takes the values 213.81 $\pm$ 14.61, 204.00, and 256.64 $\pm$ 0.37 km s$^{-1}$ \citep{elsanhouryAN,Nouh..and..Elsanhoury..20,2019ApJ...872..205L} with respective manner.

According to Eq. (\ref{sigma12}) the adopted values of dispersions $\sigma_1$, $\sigma_2$, and their ratios ($\sigma_2/\sigma_1$) are listed in Table \ref{tab5}. Considering large volume sample (i.e., Program III) into which $\sigma_2/\sigma_1\approx0.60$ and its presents a good agreement when compared with recent ones and taken into account the computed value of angular rotation rate |$A-B$| $\approx$ 26.07 $\pm$ 5.10 km s$^{-1}$ \citep{elsanhouryAN}, therefore, the adopted mean values of the Oort constant are about $A$ = 17.80 $\pm$ 0.24 km s$^{-1}$ kpc$^{-1}$ and $B$ = -9.61 $\pm$ 0.32 km s$^{-1}$ kpc$^{-1}$, these values are approximately over 80$\%$ consistency with others. 

Since OB associations include many bright stars \citep{2016MNRAS.461..794P} and high-massive ones \citep{1999AJ....117..354D} those leads us to reflect and conclude that the similarities and/or differences exit between the retrieved kinematic and rotational characteristics of Programs I, II, and III are due to volume and restrictions of gathered data as mentioned above.

%%%%%%%%%%%%%%
% Table 5 
%%%%%%%%%%%%%%
\begin{table}
\caption{Ratios of velocities dispersion and Oort's constants devoted with three Programs and recent ones.}
\centering
\tiny
\begin{tabular}{lcccc}
\hline
\textbf{Program}&\textbf{($\sigma_2/\sigma_1$)} & \textbf{$A$; km s\textsuperscript{-1} kpc\textsuperscript{-1}} & \textbf{$B$; km s\textsuperscript{-1} kpc\textsuperscript{-1}} &\textbf{References} \\  
\hline
I&0.53 & 8.74 $\pm$ 0.23 & -7.33 $\pm$ 0.20 & \text{Present study} \\
II; K-type&0.15 & 25.92 $\pm$ 0.20 & -0.15 $\pm$ 0.01 & \text{Present study} \\
II; M-type&0.14 & 25.93 $\pm$ 0.20 & -0.14 $\pm$ 0.01 & \text{Present study} \\
III; $\alpha$ Sco&0.70 & 13.29 $\pm$ 0.27 & -12.78 $\pm$ 0.27 & \text{Present study} \\
III; $\beta$ Sco&0.50 & 19.55 $\pm$ 0.23 & -6.52 $\pm$ 0.39 & \text{Present study} \\
III; $\delta$ Sco&0.75 & 11.41 $\pm$ 0.30 & -14.66 $\pm$ 0.26 & \text{Present study} \\
III; $\sigma$ Sco&0.68 & 14.01 $\pm$ 0.27 & -12.06 $\pm$ 0.28 & \text{Present study} \\
III; $\mu$ Sco&0.50 & 19.55 $\pm$ 0.23 & -6.52 $\pm$ 0.39 & \text{Present study} \\
III; $\pi$ Sco&0.50 & 19.55 $\pm$ 0.23 & -6.52 $\pm$ 0.39 & \text{Present study} \\
III; $\rho$ Oph&0.56 & 17.89 $\pm$ 0.24 & -8.18 $\pm$ 0.35 & \text{Present study} \\

inner-halo hot sd&0.80&9.38$\pm$0.34&-16.69$\pm$0.25&\text{\cite{elsanhoury24}}\\
-&0.74 & 16.06 $\pm$ 0.68 & -19.43 $\pm$ 1.37  & \text{\cite{2021AN....342..989E}} \\

K-dwarfs&- & 16.31 $\pm$ 0.61 & -11.99 $\pm$ 0.79 & \text{\cite{2021MNRAS.504..199W}} \\

M-dwarfs&0.71 & 15.60 $\pm$ 0.03 & -13.90 $\pm$ 1.80  & \text{\cite{Nouh..and..Elsanhoury..20}} \\
-&- & 15.730 $\pm$ 0.320 & -12.670 $\pm$ 0.340 & \text{\cite{2020AstL...46..370K}} \\

-&0.83&9.41$\pm$0.92&-14.24$\pm$1.38&\text{\cite{bobylev19}}\\
$d\approx$ 200 kpc&0.64&15.21$\pm$0.58&-13.83$\pm$0.89&\text{\cite{Bobylev17}}\\

\hline
\end{tabular}
\label{tab5}
\end{table}

\section{luminosity and mass functions of USC and Oph associations}
\label{sec5}

The luminosity function (LF) which is the number of stars at apparent magnitude per unit magnitude interval per unit area, allows investigators to statistically characterize the distribution of stars in the sky. All methods for determining the LF of field stars, count the number of stars as a function of their magnitudes, and several methodologies are used to estimate distances to these stars. Therefore, the absolute magnitudes and space densities could be used to estimate the LF \citep{1953stas.book.....T,1936PGro...47....1V,1966VA......7..141M}. 

The initial mass function (IMF; $\xi(log\mathfrak{M})$) represents the outcome of the star formation history and age, and all are coded in the LF profile \citep{1968MNRAS.139..221L,1993ARA&A..31..433B}.

In the present study, the classical approach of \cite{1936PGro...47....1V} is used to determine LF ($\phi(M_G)$),
\begin{equation}
    \phi(M_G)=\frac{dN}{dM_G}.
\end{equation}

This method claimed that the density of the stars is the number $dN$ of stars per cubic parsec in the magnitude interval $M_G$ to $M_G+dM_G$, and is determined by splitting the zone into shells with a specific width. A double frequency table (DFT) was created that estimates how many stars in each shell correlate to each magnitude interval per cubic parsec. Following this classical scheme of 670 USC and Oph candidates listed in Table \ref{tab3} within absolute magnitudes and distances ranging from -3.510 to 9.142 mag and 86.755 to 172.387 pc, respectively, taking into account Gaia DR does not achieve completeness at the bright end of LF ($M_G<8$) \citep{2021A&A...655A..45G}. Therefore, DFT results are displayed here in Table \ref{tab6}, where the number of members is divided into equal absolute magnitude and distance intervals. The computed LF via DFT based on the definition given by \cite{1936PGro...47....1V} is demonstrated in Table \ref{tab7} with bin size of 1.00 mag, a plot of LF appears in Figure \ref{fig4}. Clearly, LF slopes sharply by an absolute magnitude of about $M_G$ = 0.50-3.50 mag, LF has a constant density when $M_G$ ranges between 3.50 and 5.00 mag, and finally, LF posses slight depression around $M_G\sim7$ mag which may be called a Wielen dip \citep{Wielen74}.

The H-peak (or H-maximum) and R-peak (or R-maximum) in the LF are two time-dependent features \citep{1996AstL...22..466P,2004MNRAS.349.1449P} that discussed in LF of a coevally forming population of stars, where the lower mass stars are still in the pre-main sequence (PMS) phase and the more massive ones have already reached the main sequence (MS). They reported the arrival of stars on the MS is linked to the H-peak, which is situated at the brighter end of the LF, the peak associated with a shift in the slope of the mass-luminosity relation (MLR), which results in an identified peak in the stellar density at a particular luminosity. An evolution as a function of time, makes H-peak moves to the fainter end and therefore it is an indirect method for estimating the age of a region or population \citep{2021A&A...655A..45G}.

Referring to Figure \ref{fig4} for analyzing H-peak in region of ($-2.50\le M_G \le -1.50$), Wielen dip proposed by \citep{1995JKAS...28...45L,1997JKAS...30..181L} at $M_G\sim7$, and the R-peak at ($7.00\le M_G \le 9.50$), by evaluating the dispersion of whole LF and also those around H-peak, Wielen dip, and R-peak. The obtained results are about 0.674 for the whole LF, 0.024, 0.341, and 0.822 for H-peak, Wielen region, and R-peak, respectively. Therefore, the H-peak, Wielen dip, and
R-peak are smoothed out in DR3 where the dispersion around
them is either much smaller or slightly larger than whole LF.

LF is defined as the total number of stars per unit magnitude per unit area for non-van Rhijn \citep{1936PGro...47....1V} scheme. Accordingly, the LF of stars in the Solar neighborhood within 20 pc from three catalogs using the same bin size of 1.00 mag as in Wielen \cite{Wielen74}, who relied on the 1969 edition of the Gliese catalog (see \citep{Gliese1969,Gliese2015,Gliese1991}). Once more, the second reduction is within 20 pc of the Hipparcos catalog (HIP2, \cite{2007ASSL..350.....V}), and the third is built using Gaia DR2, where the peak of the LF within 20 pc is around $M_G\approx11$ mag. According to \cite{2021A&A...655A..45G} analysis using kernel density estimation (KDE), the peak at $M_G\approx6$ and the Wielen dip in the Gliese's curves of HIP2 remain present, while all features in the Gaia sample are smoothed out. The Wielen dip of the HIP2 and Gaia samples at $M_G=5$ is soothed out.

The out comes LF features showing slightly difference between these two methods based on area, volume, and distances. Therefore, these reflect on their analysis and the results as mentioned above.

%%%%%%%%%%%%%%%%%%%%%%%%%%%%%
% Table 6 
%%%%%%%%%%%%%%
\begin{table}
\caption{The DFT of USC and Oph.}
\centering
\tiny
\begin{tabular}{l|cccccccccc}
\hline
&&&&&Distance (pc)\\
Magnitude (abs.)&\begin{turn}{90} 80-90\end{turn}&\begin{turn}{90} 90-100\end{turn}&\begin{turn}{90} 100-110\end{turn}&\begin{turn}{90} 110-120\end{turn}&\begin{turn}{90} 120-130\end{turn}&\begin{turn}{90} 130-140\end{turn}&\begin{turn}{90} 140-150\end{turn}&\begin{turn}{90} 150-160\end{turn}&\begin{turn}{90} 160-170\end{turn}&\begin{turn}{90} 170-180\end{turn}\\  
\hline
%\hline
-4.00 $-$ -3.00 &&&1&&&2&&1&&\\
-3.00 $-$ -2.00 &&&&1&2&3&3&1&&\\
-2.00 $-$ -1.00 &1&&&&&2&3&2&1&\\
-1.00 $-$ 0.00  &&&3&1&&4&6&4&1&\\
0.00 $-$ 1.00   &&&2&1&2&8&2&3&&\\
1.00 $-$ 2.00   &&&2&2&2&7&9&9&2&\\
2.00 $-$ 3.00   &&&2&&1&23&29&12&2&\\
3.00 $-$ 4.00   &&&1&&3&51&49&24&9&\\
4.00 $-$ 5.00   &&&2&3&6&46&44&37&7&\\
5.00 $-$ 6.00   &&&6&4&2&31&42&33&7&2\\
6.00 $-$ 7.00   &1&&1&2&4&19&28&14&2&\\
7.00 $-$ 8.00   &&&1&&1&3&11&7&2&\\
8.00 $-$ 9.00   &&&&&&1&&&1&\\
9.00 $-$ 10.00  &&&&&&1&&&&\\

\hline
%\multicolumn{4}{l}{$^*$\cite{Hunt24}} \\   
\end{tabular}
\label{tab6}
\end{table}
%%%%%%%%%%%%%%%%%%%%%%%%%%%%%%%%%%
%%%%%%%%%%%

%%%%%%%%%%%%%%%%%%%%%%%%%%%%%
% Table 7 
%%%%%%%%%%%%%%
\begin{table}
\caption{The LF of USC and Oph utilizing \cite{1936PGro...47....1V} method.}
\centering
%\tiny
\begin{tabular}{lcccccccc}
\hline
$M_G$&-3.50&-2.50&-1.50&-0.50&0.50\\
$\log{\phi(M_G)}$&-5.716&-5.357&-5.391&-5.076&-5.081\\
\hline
\hline
$M_G$&1.50&2.50&3.50&4.50&5.50\\
$\log{\phi(M_G)}$&-4.864&-4.570&-4.271&-4.242&-4.295\\
\hline
\hline
$M_G$&6.50&7.50&8.50&9.50\\
$\log{\phi(M_G)}$&-4.535&-5.018&-6.137&-6.360\\

\hline
%\multicolumn{4}{l}{$^*$\cite{Hunt24}} \\   
\end{tabular}
\label{tab7}
\end{table}
%%%%%%%%%%%%%%%%%%%%%%%%%%%%%%%%%%
%%%%%%%%%%%

\begin{figure}
\label{fig4}
\centerline{\includegraphics[width=0.5\textwidth,clip=]{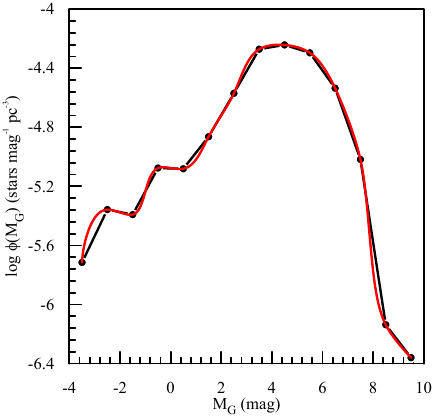}}
\caption{LF and its fitted (red solid line) of the USC and Oph associations.}
\label{fig4}
\end{figure}

The present-day mass function (PDMF) [$\phi(log\mathfrak{M})$] is defined as the number of MS stars per unit logarithmic mass interval per square parsec. The PDMF of MS field stars is related to the LF $\phi({M_G})$ of field stars by \cite{1979ApJS...41..513M} as
\begin{equation}
\phi(log\mathfrak{M})=\phi (M_{V})\bigg|\frac{dM_{V}}{dlog\mathfrak{M}}\bigg|2H(M_{V})f(M_{V}).
\label{phi}
\end{equation}

To produce PDMF, the visual ($V$) magnitude with the provided ($G - V$) color index to the $G$-band magnitude in DR3 according to the recipe in \cite{2021A&A...649A...3R}, so that PDFM is in $M_G$. LF is converted to MF by $dM_{V}/dlog\mathfrak{M}$ which is the scale with the absolute value of the derivative of the MLR \citep{2021A&A...655A..45G}. When LF is integrated perpendicular to the Galaxy's plane, assuming an exponential distribution with scale height $H(M_{G})$, the result is the term $2H (M_{G})$. According to \cite{1979ApJS...41..513M}, the fraction of objects at a particular magnitude is indicated by the factor $f(M_{G})$. The MS today (i.e., $T_{0}$) will contain stars with MS lifetimes ($T_{ms}$) larger than the Galaxy's age, independent of their formation time. These stars have an identical PDMF [$\phi(log\mathfrak{M})$] and IMF [$\xi(log\mathfrak{M})$] \citep{2024ApJ...966..169P,2018A&A...620A..39J,1997JKAS...30..181L,1979ApJS...41..513M}:

\begin{equation}
\phi(log\mathfrak{M})=\xi(log\mathfrak{M}),~~~T_{ms}\ge T_{0}.
\end{equation}

To compute IMF of USC and Oph associations, referring to Eq. (\ref{phi}) with adopted relationship \citep{1979ApJS...41..513M} the obtained results are drawn here in Table \ref{tab8} and the retrieved plot in Figure \ref{fig5}. For analyzing H-peak in regions of ($6.053\ge\frac{M}{M_{\odot}}\ge 4.188$), Wielen dip proposed by \citep{1995JKAS...28...45L,1997JKAS...30..181L} at $M\simeq0.586M_{\odot}$, and the R-peak within ($0.586\ge\frac{M}{M_{\odot}}\ge 0.372$) by evaluating the dispersion of whole IMF and also those around H-peak, Wielen dip, and R-peak. Therefore, the dispersion are about 1.541 for the whole IMF, 0.486, 0.076, and 0.060 for H-peak, Wielen region, and R-peak, respectively. Clearly, H $\&$ R peaks and Wielen dip are smoothed out in DR3 where dispersion around all remains much smaller than those obtained for the whole IMF. This confirmation and conclusion is compatible with those reported in \cite{2011ISRAA2011E...1E} and recently with Gaia DR2 \citep{2021A&A...655A..45G}.

\begin{figure}
\label{fig5}
\centerline{\includegraphics[width=0.5\textwidth,clip=]{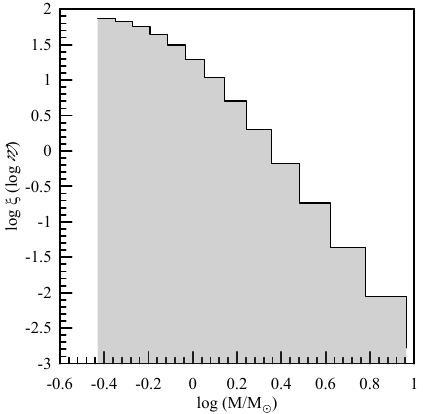}}
\caption{The IMF of the present study of the USC and Oph associations.}
\label{fig5}
\end{figure}

%%%%%%%%%%%%%%
% Table 8 
%%%%%%%%%%%%%%
\begin{table}
\caption{The USC and Oph associations IMF.}
\centering
%\tiny
\begin{tabular}{lcc}
\hline
{\bf $M_G$}&{\bf $\log(M/M_{\odot})$}&{\bf $log\xi~(log\mathfrak{M})$}\\   
\hline
-3.50&0.964&-2.779\\
-2.50&0.782&-2.051\\
-1.50&0.622&-1.363\\
-0.50&0.481&-0.735\\
0.50&0.356&-0.178\\
1.50&0.244&0.303\\
2.50&0.144&0.706\\
3.50&0.052&1.035\\
4.50&-0.033&1.295\\
5.50&-0.115&1.495\\
6.50&-0.193&1.644\\
7.50&-0.271&1.751\\
8.50&-0.349&1.824\\
9.50&-0.430&1.870\\

\hline
\end{tabular}
\label{tab8}
\end{table}
%%%%%%%%%%%
%%%%%%%%%%%

\section{discussion and conclusion}
\label{sec6}
For adopted three Programs of Upper Scorpius associations. The computed kinematics, velocity ellipsoid, Solar elements, convergent points, Oort's constants, and luminosity and mass functions are studied and analyzed using the most precise three-dimensional map of our Galaxy provided by Gaia DR3.

For the first Program, the row data of $U,~V,$ and $V$ are extracted within \cite{2018MNRAS.477L..50G} for kinematic analysis purposes and then updated using DR3 to get about 51 Upper Scorpius stars as listed in Table \ref{tab1}. Program II was developed using DR2 \citep{2018A&A...616A...1G} and new spectroscopy members of Upper Scorpius \citep{2018AJ....156...76L}. Moreover, the retrieved data are also updated with DR3 to get about two subgroups; K-type (59 stars) and M-type (83 stars) as numerated in Table \ref{tab2}. Third Program is devoted to seven groups \citep{2022A&A...667A.163M} in the USC and Oph associations, and the retrieved data are about 670 stars with Gaia DR3 as seen in Table \ref{tab3}.

Developing our code to compute kinematical parameters and the obtained conclusions are drawn here in Table \ref{tab4} for Programs I, II, and III and their subgroups, into which mean space velocities ($\overline{V_x}$, $\overline{V_y}$, $\overline{W_z}$; km s$^{-1}$) and spatial velocities ($\overline{U}$, $\overline{V}$, $\overline{W}$; km s$^{-1}$) due to equatorial and Galactic coordinates, respectively are presented, velocities dispersion ($\sigma_1$, $\sigma_2$, $\sigma_3$; km s$^{-1}$), Solar elements ($S_{\odot}$; km s$^{-1}$) $\&$ ($l^{o}_A$, $b^{o}_A$, $\alpha^{o}_A$, $\delta^{o}_A$), and the convergent points ($A_o$, $D_o$)$^{o}$. Table \ref{tab5} gives a comparison of the present study and other recent ones of ratios ($\sigma_2/\sigma_1$) and Oort's constants ($A$, $B$; km s$^{-1}$ kpc$^{-1}$).

The density distribution in a specific direction of the Galactic plane may be deduced by the total number of stars per unit volume between absolute magnitudes $M_G$ and $M_G + dM_G$, well-defined as luminosity function \citep{1936PGro...47....1V}. On the other hand, the ability to turn the luminosity function into the mass function is particularly significant, and the related initial mass functions may be analyzed due to the simple integration of evolutionary effects.

The following conclusions are obtained:
\begin{itemize}
   \item Computed mean spatial velocities ($\overline{U},~\overline{V}$, $\overline{W}$; km s$^{-1}$) for USC as follows; -7.21 $\pm$ 2.69, -16.17 $\pm$ 4.02, -7.29 $\pm$ 2.70 (Program I), -5.25 $\pm$ 2.29, -15.98 $\pm$ 4.00, -7.42 $\pm$ 2.72 (Program II), and -5.05 $\pm$ 2.25, -16.27 $\pm$ 4.03, -7.23 $\pm$ 2.69 (Program III). These values are in line with those computed ones (e.g., \cite{2018MNRAS.477L..50G}, \cite{2022A&A...667A.163M}, \citep{2020AJ....160...44L}, and \cite{2018MNRAS.476..381W}).

    \item Mean velocities dispersion ($\overline{\sigma_1},~\overline{\sigma_2},~\overline{\sigma_3}$; km s$^{-1}$) for the three Programs are as follows; 19.60 $\pm$ 4.43, 10.41 $\pm$ 3.23, 1.63 $\pm$ 0.78 (Program I), 18.30 $\pm$ 4.23, 2.60 $\pm$ 0.62, 1.17 $\pm$ 0.93 (Program II), and 1.36 $\pm$ 0.02, 0.80 $\pm$ 0.01, 0.96 $\pm$ 0.01 (Program IIII).
    
    \item Here, the mean Solar velocities ($\overline{S_\odot}$; km s$^{-1}$) are about 19.14 $\pm$ 4.37, 18.39 $\pm$ 4.29, and 18.62 $\pm$ 4.32 in unite (km s$^{-1}$) of Program I, II, and III, respectively. Other Solar elements ($l_A$, $b_A$, $\alpha_A$, $\delta_A$) numerically are listed in Table \ref{tab4}.

    \item Convergent points ($A_o,~D_o$)$^{o}$ are as follows; 94$^o$.21 $\pm$ 0$^o$.11, -38$^o$.43 $\pm$ 0$^o$.02 (Program I), 94$^o$.35 $\pm$ 0$^o$.11, -46$^o$.50 $\pm$ 0$^o$.15 (Program II), and 95$^o$.91 $\pm$ 0$^o$.10, -44$^o$.42 $\pm$ 0$^o$.15 (Program III). These obtained results are confirmed with recent ones (e.g., \cite{1999MNRAS.306..381D,2018MNRAS.477L..50G}) and \cite{2018MNRAS.476..381W} by Jones method \citep{1971MNRAS.152..231J}.
    
   \item Computation the rotational constants are based on velocities dispersion ratio ($\sigma_2/\sigma_1$) and the devoted results are appears in Table \ref{tab5}. Here, the adopted mean values of the Oort constant are about $A$ = 17.80 $\pm$ 0.24 km s$^{-1}$ kpc$^{-1}$ and $B$ = -9.61 $\pm$ 0.32 km s$^{-1}$ kpc$^{-1}$. These values are consistency with other results. Notably, the selection of kinematic model characteristics may be influenced by the over or underestimation of $A$, $B$, and other dependent parameters \citep{1990MNRAS.244..247L,1987AJ.....94..409H}. A detailed summary of all kinematical structure parameters is presented in Table \ref{tab5}.

  \item Hunting DR3 data of USC and Oph associations and their distribution to examine their luminosity function \citep{1936PGro...47....1V} and its relation to the initial mass functions. Concluded that the obtained dispersion around three regions for H-peak, Wielen dip, and R-peak are either smaller or larger than those dispersion around whole luminosity and initial mass functions, This makes us and others on the side of an absence of these regions in OB young associations. These obtained results are compatible and confirmed by \cite{2011ISRAA2011E...1E} and in Gaia DR2 \citep{2021A&A...655A..45G}. 
\end{itemize}

\section*{Acknowledgments}

We sincerely thank the anonymous referee for their valuable suggestions, which greatly enhanced the quality of this paper. This work presents results from the European Space Agency space mission Gaia. Gaia data are being processed by the Gaia Data Processing and Analysis Consortium (DPAC). Funding for the DPAC is provided by national institutions, in particular, the institutions participating in the Gaia MultiLateral Agreement (MLA). The Gaia mission website is \url{https://www.cosmos.esa.int/gaia}. The Gaia archive website is \url{https://archives.esac.esa.int/gaia}. The author extend their appreciation to the Deanship of Scientific Research at Northern Border University, Arar, KSA for funding this research work through the project number "NBU-FFR-2025-237-02”.

\subsection*{Conflict of interest}

The authors declare no financial or commercial conflict of interest.

\subsection*{ORCID}
\textit{Waleed H. Elsanhoury} \url{https://orcid.org/0000-0002-2298-4026}

\bibliography{Wiley-ASNA}%

\section*{Author Biography}

\begin{biography}{\includegraphics[width=60pt,height=70pt,draft]{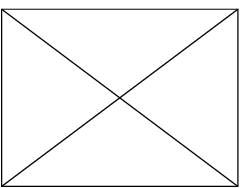}}{\textbf{Waleed H. Elsanhoury.} born in 1972 in Egypt, is a full professor at the National Research Institute of Astronomy and Geophysics (NRIAG), located in Helwan, Cairo, Egypt. He is currently affiliated with the College of Science, Physics Department, at Northern Border University in Saudi Arabia. He earned his Ph.D. in Theoretical Physics in 2009 from Physics Department, Faculty of Science, at Al-Azhar University, Egypt. Prior to that, he received his M.Sc. in Physics in 2004 from Physics Department, Faculty of Science, at Helwan University, Egypt, and his B.Sc. in Physics (Special Degree) in 1994 from Physics Department, Faculty of Science, at Alexandria University, Egypt.
His research interests focus on computational and mathematical physics, with particular emphasis on their applications in astronomy and astrophysical models, utilizing the Mathematica software package. His current research is dedicated to investigating the photometry, astrometry, dynamical evolution, and kinematic behaviors of open star clusters and their associations in the Gaia era. His body of work includes 53 publications, which have been cited approximately 178 times.}
\end{biography}

\end{document}